\documentclass[sigconf]{acmart}

\usepackage{bbding}
\usepackage{graphicx}
\usepackage{geometry}
\usepackage{subfigure}
\usepackage{makecell}
\usepackage{xcolor}
\usepackage{subcaption}
\usepackage{hyperref} 
\usepackage{soul} 
\usepackage{xcolor}

\copyrightyear{2026}
\acmYear{2026}
\setcopyright{cc}
\setcctype{by}
\acmConference[CHI '26]{Proceedings of the 2026 CHI Conference on Human Factors in Computing Systems}{April 13--17, 2026}{Barcelona, Spain}
\acmBooktitle{Proceedings of the 2026 CHI Conference on Human Factors in Computing Systems (CHI '26), April 13--17, 2026, Barcelona, Spain}
\acmDOI{10.1145/3772318.3791088}
\acmISBN{979-8-4007-2278-3/2026/04}

\begin{document}

\title{Reimagining Legal Fact Verification with GenAI: Toward Effective Human-AI Collaboration}

\author{Sirui Han}
\authornote{Both authors contributed equally to this work.}
\affiliation{%
  \institution{The Hong Kong University of Science and Technology}
  \city{Hong Kong SAR}
  \country{China}}
\email{siruihan@ust.hk}

\author{Yuyao Zhang}
\authornotemark[1]
\affiliation{%
  \institution{The Hong Kong University of Science and Technology}
  \city{Hong Kong SAR}
  \country{China}}
\email{yzhang075@connect.ust.hk}

\author{Yidan Huang}
\affiliation{%
  \institution{The Hong Kong University of Science and Technology}
  \city{Hong Kong SAR}
  \country{China}}
\email{yhuangjr@connect.ust.hk}

\author{Xueyan Li}
\affiliation{%
  \institution{The Hong Kong University of Science and Technology}
  \city{Hong Kong SAR}
  \country{China}}
\email{xueyanli0329@gmail.com}

\author{Chengzhong Liu}
\affiliation{%
  \institution{The Hong Kong University of Science and Technology}
  \city{Hong Kong SAR}
  \country{China}}
\email{chengzhong.liu@connect.ust.hk}

\author{Yike Guo}
\authornote{Corresponding author.}
\affiliation{%
  \institution{The Hong Kong University of Science and Technology}
  \city{Hong Kong SAR}
  \country{China}}
\email{yikeguo@ust.hk}

\begin{abstract}
Fact verification is a critical yet underexplored component of non-litigation legal practice. 
While existing research has examined automation in legal workflow and human-AI collaboration in high-stakes domains, little is known about how GenAI can support fact verification, a task that demands prudent judgment and strict accountability. 
To address this, we conducted semi-structured interviews with 18 lawyers to understand their current verification practices, attitudes toward GenAI adoption, and expectations for future systems. 
We found that while lawyers use GenAI for low-risk tasks like drafting and language optimization, concerns over accuracy, confidentiality, and liability are currently limiting its adoption for fact verification.
These concerns translate into core design requirements for AI systems that are trustworthy and accountable.
Based on these, we contribute design insights for human-AI collaboration in legal fact verification, emphasizing the development of auditable systems that balance efficiency with professional judgment and uphold ethical and legal accountability in high-stakes practice.
\end{abstract}

\ccsdesc[500]{Human-centered computing~Empirical studies in HCI}

\keywords{Legal fact verification, GenAI, Workflow}

\begin{teaserfigure}
  \includegraphics[width=\textwidth]{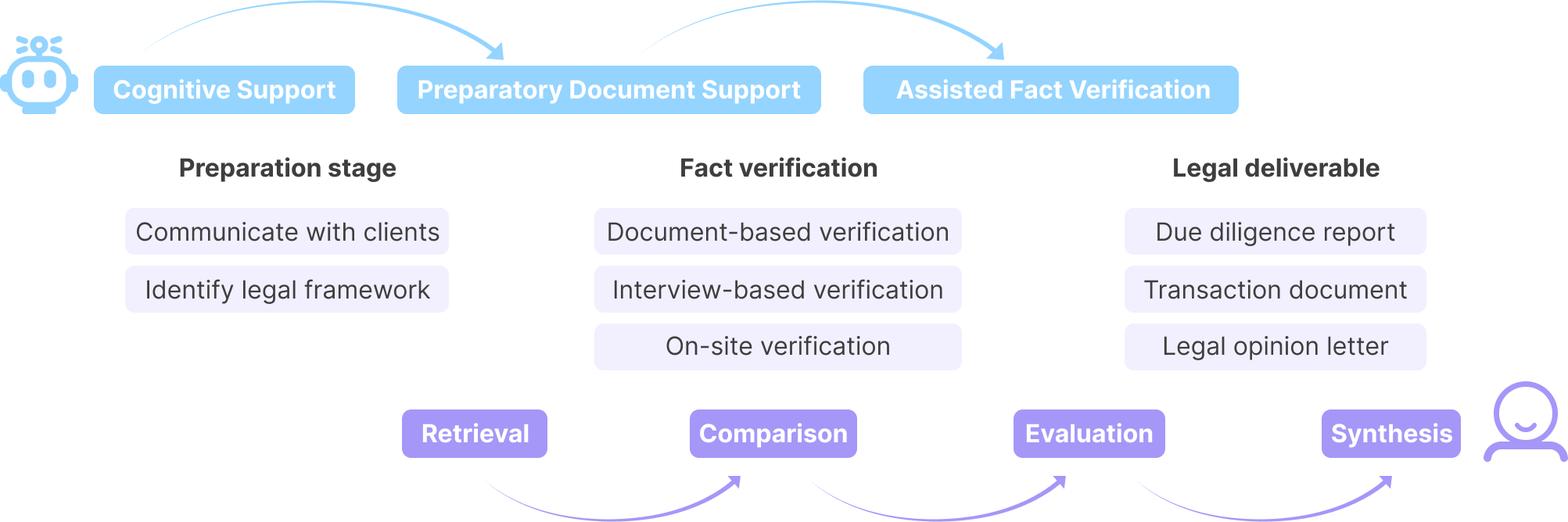}
  \caption{Legal Fact Verification Workflow with GenAI. Legal fact verification operates as a dynamic sensemaking activity where humans combine retrieval, comparison, evaluation, and synthesis to construct a coherent, defensible narrative of the matter, while GenAI provides targeted support. It offers cognitive aid for client communication and legal framework identification in the preparation stage, assists with preparatory document work during fact verification, and contributes to generating due diligence reports, transaction documents and legal opinion letters for legal deliverables.}
  \Description{.}
  \label{fig:teaser}
\end{teaserfigure}

\maketitle

\section{Introduction}
Fact verification is a critical process in non-litigation \footnote{Litigation refers to dispute resolution through formal judicial proceedings. In contrast, non-litigation refers to legal work outside courts, such as transactions, compliance, and advisory services.} legal practice, ensuring that all information supporting contractual commitments, regulatory filings and advisory decisions is not only accurate and complete, but also consistent and defensible \cite{schwarcz2006explaining, michels2010internal, stankovic2023legal, spreeuwenberg2001role}. 
High-stakes business transactions \cite{mann2002business} and compliance processes \cite{benedek2024compliance} depend on the careful verification of factual information.
A single oversight, such as failing to detect an undisclosed beneficial owner or misinterpreting a cross-border regulatory requirement, can lead to severe consequences including financial penalties, reputational harm, and prolonged disputes \cite{fifield2010three}.

However, verification is far from a straightforward checklist.
It is a complex and interpretive process that requires lawyers to synthesize a large amount of information from diverse sources \cite{rowe1999legal}, such as client-provided materials, official registries, and third-party reports, while reconciling inconsistencies and evaluating the credibility of each source \cite{morris1941law}.
Moreover, as new documents and contextual details emerge during the course of legal services, lawyers revisit and refine their factual understanding.
Consequently, rather than a static confirmation of individual facts, fact verification operates as a dynamic sensemaking activity that combines retrieval, comparison, evaluation, and synthesis to construct a coherent and defensible narrative of the matter \cite{chesler2018telling}. 
Such tasks are cognitively demanding and operationally fragmented.
Lawyers must reconcile voluminous, often contradictory information under significant time constraints \cite{hopkins1999cross}. 

Existing legal technology solutions \cite{park2021survey, dudchenko2023legal, rodgers2023technology} provide partial relief by automating specific subtasks \cite{myroslavskyi2025role}.
For example, Westlaw \cite{westlaw} facilitates legal research across jurisdictions, and analytics tools such as Luminance \cite{luminance} accelerate due diligence by detecting clause-level inconsistencies.
However, these tools operate within structured, task-specific contexts and fall short in supporting cross-document reasoning or interpretive synthesis.
They can highlight a missing warranty clause in a contract, but they cannot determine that an ownership disclosure in one agreement conflicts with shareholder information contained in another document.
Such limitations leave lawyers responsible for the interpretive work required to maintain factual coherence across dynamic and distributed information landscapes \cite{bues2016legaltech}.

Currently, Generative AI (GenAI) is being explored for implementation in professional fields such as medicine \cite{yim2024preliminary, health2023embracing}, finance \cite{shabsigh2023generative, joshi2025review}, and scientific research \cite{andersen2025generative, hanafi2025generative} through diverse technical approaches, including fine-tuning large language models (LLMs) \cite{naveed2025comprehensive, thirunavukarasu2023large}, integrating domain-specific knowledge retrieval \cite{lu2024empowering}, and optimizing task instruction design \cite{zhou2024star}. 
Unlike traditional automation, GenAI with advanced capabilities in language understanding, contextual reasoning, and multi-turn interaction,  can potentially assist fact verification \cite{dwivedi2025potential, ajevski2023chatgpt}.
However, its application to fact verification, the most judgment-dependent and high-stakes aspect of non-litigation practice, remains largely unexplored. Integrating GenAI introduces critical Human-Computer Interaction (HCI) questions: How should GenAI engage in fact verification without undermining professional accountability? How can interaction designs support interpretive flexibility, sensemaking, and trust while mitigating over-reliance? To address this gap, we conducted semi-structured interviews with 18 experienced non-litigation lawyers. Specifically, we focus on the following research questions:

\textbf{RQ1: How do lawyers conduct fact verification in non-litigation practice, and where does GenAI fit in these workflows?}

\textbf{RQ2: What cognitive and operational challenges influence lawyers’ willingness and ability to use GenAI for fact verification?}

\textbf{RQ3: Given these work practices and expectations, what forms of human-AI interaction can best support reliable and accountable fact verification?}

By examining these questions, our study offers an integrated view of how lawyers navigate the ongoing construction of factual understanding and where current tools fall short in supporting this work. The interview findings illuminate the practical realities behind fact verification by showing how lawyers synthesize evolving information, reconcile inconsistencies, and maintain defensibility under time pressure. This reveals how GenAI might meaningfully contribute to or potentially complicate the lawyers’ efforts to improve their work efficiency. In doing so, this research advances a deeper understanding of what reliable and accountable fact verification requires in practice, and outlines how future HCI can better support the interpretive and judgment-dependent nature of non-litigation legal work.
\section{Related Work}
\subsection{Legal Technology for Fact Verification}
Legal technology, or LegalTech for short, refers to tools and systems that leverage technologies such as artificial intelligence (AI), big data, and cloud computing to optimize legal business processes \cite{webb2022legal, armour2020ai, hongdao2019legal}. Its core objective is to improve the efficiency of legal work, reduce costs, and minimize human errors through technical means, covering the entire business chain, from contract management and case retrieval to compliance review \cite{veith2016legal}. 
Early systems primarily focused on single-task automation, such as clause comparison, semantic search for statutes, and document drafting.
The focus on the automation of individual tasks has yielded tangible results in various legal domains \cite{kabir2023iot}. 
In the field of contracts, previous works show that combining information design and computer codification can enhance communication, participation, and usefulness across the entire life-cycle of contracting \cite{barton2019successful}.
Kozlova et al. \cite{kozlova2020smart} elaborated the advantages and disadvantages of using both smart contracts and classic contracts in contractual practice. Their research suggests that standardized contracts can be completely replaced by smart contracts.
Within retrieval and analysis, intelligent systems enable semantic matching for case and regulation searches, recommend similar cases, and analyze judicial trends of such cases \cite{moens2001innovative, pietrosanti1999advanced}. 
For example, Nguyen et al. \cite{nguyen2024attentive} developed Attentive Convolutional Neural Network (Attentive CNN) and Paraformer, using deep neural networks with attention mechanisms. These tools proved better performance in terms of retrieval performance across datasets and languages. 
For documents and processes, automation extends to filling legal documents, such as complaints and evidence lists, as well as handling repetitive tasks such as format review and information extraction \cite{di2024streamlining, singireddy2024ai}. 
This single-point automation provides significant value by reducing the mechanical labor of lawyers on low-value tasks, such as reviewing standardized contract clauses \cite{iyelolu2024legal}.

Fact verification remains inherently dependent on nuanced professional judgment, as it requires contextual understanding, and legal sufficiency evaluation. However, most LegalTech tools are still confined to mechanical or highly structured subtasks, leaving complex, judgment-dependent activities such as fact verification to human professionals \cite{pasquale2019rule}.
For example, while AI tools can detect missing data or flag inconsistent entries in contracts, they lack the capability to evaluate the factual accuracy of complex disclosures, making them insufficient for tasks requiring legal judgment and accountability \cite{remus2017can, mcginnis2013great}.
 This gap motivates investigation into whether and how GenAI can assist verification without undermining professional standards.

\subsection{GenAI for Legal Practice}
GenAI is being actively explored for various applications within the legal domain, such as document drafting, legal research, and contract analysis, significantly enhancing efficiency, accessibility, and the overall quality of legal processes \cite{wondracek2025generative}.
In document drafting, AI-powered tools like the one developed by Luzi et al. \cite{de2023cicero} assist in generating civil judgments, while legal writing aids such as LegalWriter \cite{weber2024legalwriter} provide personalized feedback to law students, leading to more structured and persuasive case solutions.
In addition, studies have found that not only are AI-generated legal summaries more accessible to non-experts \cite{ash2024translating}, but laypeople also exhibit a marked preference for legal advice from LLMs, even when they can distinguish them from human lawyers \cite{schneiders2025objection}.
These studies collectively indicate that AI-powered tools are not only enhancing the clarity and accessibility of legal information for the general public but are also proving effective in improving the quality and persuasiveness of legal writing among students and professionals.
In legal research and information retrieval, LLMs use semantic understanding and generation capabilities to transform traditional keyword-based search into a knowledge-based question-answering system
can enhance the efficiency and user experience of legal retrieval \cite{contini2024unboxing, schwarcz2025ai, shilaskar2025genai, halgin2024intelligent, trippas2024adapting}.
This facilitates precise extraction of relevant information from vast documents, ensuring context continuity for lawyers during reviews \cite{geetha2024conversational}.
Techniques combining knowledge graphs with document analysis further enhance the accuracy of understanding \cite{sumukh2024framework}.
Beyond practice, research in the field of legal education indicates that well-designed evaluation  systems incorporating GenAI can enhance the academic performance of law students \cite{head2024assessing, alimardani2024generative, g2024integrating}.

The ability of LLMs to comprehend legal language and perform legal reasoning primarily depends on the accuracy of their output. High accuracy is crucial for mitigating the hallucinatory effect \cite{strkak2024generative}.
While studies indicate GenAI's potential to streamline court operations \cite{saleem2024use, shi2025legalreasoner}, its benefits are tempered by significant concerns regarding biased outputs, a lack of explainability, and ambiguous accountability \cite{wondracek2025generative, nidhisree2024generative, walkowiak2024generative}.
These challenges necessitate lawyers a responsible approach to integration \cite{pierce2024lawyers}, emphasizing human oversight and adherence to ethical standards \cite{white2025integration, wright2024professionals}.
From a regulatory standpoint, frameworks have been proposed to ensure accountability through human oversight, clear role delineation, and improved AI literacy \cite{carnat2024addressing}.
This body of discourse highlights GenAI’s profound potential across the legal ecosystem, but also underscores that its value is contingent on the establishment of trustworthy and fair application mechanisms \cite{de2024artificial, charlotin2025genai, cao2025safelawbench}.

Recent studies have examined how GenAI is being adopted across legal practice, particularly in activities such as drafting, research, information retrieval, and legal education. While this body of work demonstrates substantial potential for improving efficiency and knowledge access, few studies shift the focus from system capabilities to the lived, situated practices of lawyers. The theme how GenAI use intersects with the accountability demands and risk structures of legal practice connect directly to broader discussions of human–AI collaboration in high-stakes work, which we discuss next.

\subsection{Human-AI Collaboration in  High-stakes Work}
Human-AI collaboration has become a central theme in HCI research, especially in high-stakes professional fields like law \cite{delgado2022uncommon, bertrand2024ai, constantinides2024implications}, healthcare \cite{park2019identifying}, and finance \cite{sachan2024human}.
In these domains, AI is deployed not to replace human judgment but to augment professional expertise by enhancing productivity, improving decision-making, and reducing errors. 
For instance, Xi et al. \cite{xi2024measuring} demonstrated that human-AI collaborative systems can significantly shorten legal review cycles and improve error detection rates in legal contract review.

However, these efficiency gains are accompanied by critical challenges concerning trust, transparency, and accountability. Vallayil et al. \cite{vallayil2023explainability} highlighted explainability as an essential requirement for modern AI systems.
Jin et al. \cite{jin2024beyond} examined how public defenders interact with government-deployed computational forensic software, highlighting the difficulties legal practitioners encounter when scrutinizing automated decision-making tools.
Herrewijnen et al. \cite{herrewijnen2024requirements} explored law enforcement domain experts' needs for explainable AI (XAI) in their daily work.
They advised that in such a high-stakes domain, AI systems should align the machine learning model's capabilities with users’ domain knowledge to improve usability.
Guo et al. \cite{guo2025live} believed user trust in the AI's reliability remained a critical factor, particularly for high-stakes scenarios.
Wang et al. \cite{wang2019designing} proposed a framework that connects human reasoning with XAI design, providing developers with a practical pathway for building more effective explainable user interfaces.

Beyond explainability, high-stakes domains including media \cite{guo2022survey, du2022synthetic}, medicine and finance \cite{ali2023explainable} increasingly demand robust approaches for verifying factual correctness.
In legal work specifically, this demand is compounded by strict requirements for anonymization and privacy protection, as AI-assisted verification involves processing sensitive personal information. Recent research on Zero-Knowledge Proofs (ZKPs) demonstrates how cryptographic methods can validate sensitive information without revealing the underlying data \cite{bamberger2022verification}. However, while such cryptographic solutions strengthen privacy, they do not resolve the core judgmental challenge, which is that they cannot evaluate whether the data itself is factually accurate, contextually appropriate, or legally sufficient.

In sum, existing literature underscores that effective human–AI collaboration in high-stakes domains requires systems designed to augment rather than replace professional judgment, while ensuring trust, transparency, and privacy. However, current studies rarely examine how GenAI can be integrated into workflows that fundamentally rely on human oversight, particularly in legal fact verification. Our research builds on this by exploring how GenAI can assist in legal fact verification processes, enhancing efficiency while safeguarding human accountability and professional discretion.

\section{Methodology}

\subsection{Participants and Recruitment}
We recruited 18 non-litigation lawyers, each with at least one year of professional experience in non-litigation practice. Lawyers with less than one year of experience and those whose primary work involved litigation were excluded from the study. Recruitment was carried out through snowball sampling. The 3rd author shared a recruitment message through personal social media, and initial respondents invited colleagues from their professional networks to participate. The final sample included 7 male and 11 female lawyers who worked in practice areas such as corporate transactions, cross-border compliance, and contract review. We intentionally included participants without prior experience in GenAI in order to capture a wide range of verification practices, existing challenges, and expectations regarding the role of GenAI in legal fact verification. Each participant received a 200 RMB JD E-gift card as compensation for their time. Table~\ref{p} details the characteristics of the participants.

\begin{table*}[h]
  \caption{Summary of Participants’ Characteristics. Abbreviations used in this table: LC = Legal Counsel; PE = Private Equity; IPO = Initial Public Offering; I\&F = Investment and Financing; M\&A = Mergers and Acquisitions.}
  \label{p}
  \begin{tabular}{ccclc}
    \toprule
    \textbf{ID} & \textbf{Gender} & \textbf{Yrs of Exp.} & \textbf{Key Projects} & \textbf{Self-reported Use of GenAI} \\
    \midrule
    P1 & Male& 10 years & LC for government and enterprise & Moderate\\
    P2& Female & 2 years & PE Investment, IPO & Very frequent\\
    P3& Male& 1.5 years & Cross-border I\&F, M\&A & Frequent\\
    P4& Male& 6 years & Financing, bankruptcy & Very frequent\\
    P5& Male& 2 years & LC for capital market & Moderate\\
    P6& Female& 1 year & Long-term LC & Moderate\\
    P7& Female& 6 years & I\&F & Moderate\\
    P8& Male& 2 years & IPO & Never used\\
    P9& Male& 1 year & Due diligence & Occasional\\
    P10& Female& 3 years & Compliance services for private funds  & Very frequent\\
    P11& Male& 3.5 years & Long-term LC & Very frequent\\
    P12& Female& 1 year & Long-term LC & Frequent\\
    P13& Female& 1.5 years & Company M\&A & Frequent\\
    P14& Female& 1 year & Contract review, document writing & Very frequent\\
    P15& Female& 1.5 years & Compliance review & Very frequent\\
    P16& Female& 1.5 years & Cross-border I\&F & Moderate\\
    P17& Female& 1 year & IPO, bankruptcy & Very frequent\\
    P18& Female& 1 year & Contract review and drafting, document writing & Frequent\\
    \bottomrule
  \end{tabular}
\end{table*}

\subsection{Data Collection}
We conducted semi-structured interviews remotely using Tencent Meeting to provide flexibility for participants’ schedules. Prior to participation, they were informed about the purpose of the study, the voluntary nature of participation, and how their data would be handled. They were also assured that they could withdraw at any time without penalty. Each interview lasted approximately 45 to 60 minutes and was conducted by the same author to maintain consistency and ensure familiarity with the entire dataset. All sessions were audio-recorded with permission and automatically transcribed using Tencent Meeting’s transcription service. The research team manually reviewed and corrected transcription errors to ensure accuracy.

The interview questions were designed to align with our research questions while allowing participants to elaborate on their experiences in their own terms. The first part focused on fact verification in practice, probing how participants defined verification, what steps they followed, and what cognitive or operational challenges they encountered. The second part examined lawyers’ perceptions of GenAI, including their awareness of such technologies, perceived opportunities and risks, and scenarios where GenAI could potentially support verification tasks. The final part invited participants to envision collaboration with GenAI, exploring preferred interaction modalities, levels of control, and expectations for trust and reliability. The full list of questions is provided in the supplementary materials.

\subsection{Data Analysis}
Two authors applied thematic analysis following Braun and Clarke’s six-phase framework \cite{braun2006using}, with all identifying details anonymized due to the sensitivity of legal work. Both authors began with the familiarization phase, reading the transcripts multiple times to gain a holistic understanding of the data.
During the initial coding phase, each author independently coded the transcripts in a shared Microsoft Excel sheet organized by participant ID, interview prompt, verbatim excerpt, descriptive code, and analytic memo. We applied descriptive codes to segments relevant to the research questions and added new codes whenever excerpts did not align with existing ones. Both coders progressed in participant order and occasionally revisited earlier transcripts when later excerpts suggested that previous coding required refinement. After roughly every 4 interviews, we compared our coding decisions, discussed ambiguous excerpts, clarified overlapping labels, and resolved disagreements by returning to transcript context and consulting memos, which enabled us to maintain consistent coding rules throughout the process. This process was inductive, allowing codes to form from the data rather than being predetermined.

Once initial coding was complete, we began grouping related codes into broader analytic categories, such as “document generation with GenAI,” “human verification and ultimate accountability,” and “external constraints from client.” Building on these categories, we moved into the stage of searching for themes by reorganizing clusters into early thematic groupings and exploring different configurations of how the categories might relate to each other. This involved repeatedly mapping and remapping the relationships among categories to identify potential overarching concepts.
We then iteratively refined these preliminary groupings into candidate themes by evaluating their internal coherence and their ability to represent recurring meanings in the dataset. For example, a candidate theme such as “conditional reliance on GenAI in legal verification” might be refined from initial categories like “AI-assisted drafting for efficiency” and “mandatory human verification for accountability.” Each candidate theme was compared against the full corpus to ensure that it reflected the data comprehensively rather than selectively. During this process, we looked for negative or contradictory cases, adjusted boundaries when a code appeared to span more than one interpretive space, and merged categories that were overly narrow or conceptually overlapping. Revisions continued until the themes provided a stable and coherent representation of participants’ practices and perspectives related to legal fact verification. The final phase involved naming the themes and selecting representative excerpts that best captured each theme. All quotations were paraphrased or lightly edited for clarity without altering meaning.

The research team’s positionality influenced the analysis in important ways. The 2nd author, trained in HCI, conducted all interviews. This background as an outsider to the legal domain reduced assumptions of shared expertise, encouraging participants to elaborate on their workflows and explain domain-specific reasoning in detail. At the same time, another author with professional legal experience contributed to the coding process and thematic refinement to address the lack of direct legal experience that may have shaped how some concepts were initially interpreted. This author provided contextual insights into legal terminology, fact verification practices, and organizational norms, which ensured that interpretations accurately reflected the realities of legal work rather than being abstracted away from practice. Disagreements between the two authors were resolved through open discussion until consensus was reached, with each perspective informing the final themes.

\section{Findings}
Based on our research questions, our findings center around three key aspects from the interviews: (1) current practices, (2) challenges shaping GenAI adoption and (3) envisioning future human-AI collaboration.

\subsection{Current Practices}
To understand RQ1, we structured our analysis around three themes: (1) the verification workflow, (2) the complexity of fact verification and (3) GenAI can reduce uncertainty and burden in fact verification.

\subsubsection{Workflow of Fact Verification.} 
\label{4.1.1}
\begin{figure}[h]
    \centering   \includegraphics[width=\linewidth]{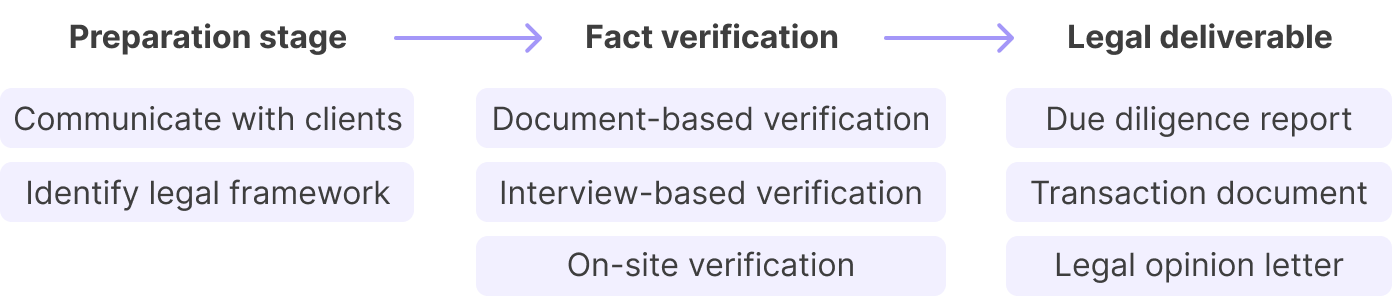}
    \caption{A Common Fact Verification Workflow Summarized from 18 Participants.}
    \label{workflow}
\end{figure}

Fact verification is an integral part of non-litigation legal practice, underpinning a wide range of work including contract review, due diligence, cross-border compliance, and advisory services. Through interviews with 18 lawyers, we identified a common workflow that begins with defining the legal and business context of the task, as visually outlined in Figure \ref{workflow}. Lawyers emphasized that verification is embedded throughout the workflow rather than being a standalone step, guiding decisions from document preparation to final recommendations. Among the participants, 15 out of 18 reported that an initial understanding of client objectives and applicable legal frameworks is essential before any verification work, as this baseline shapes the scope, sequencing, and priority of subsequent tasks.

Lawyers draw on diverse sources for fact verification, including legal provisions, internal company documents, contracts, case law, and third-party public information. Document-based verification constitutes the most frequent and foundational form, particularly in routine advisory and transactional work. Lawyers use official platforms (e.g., Tianyancha, Qichacha) to confirm corporate registration, shareholder structure, and litigation history, while cross-checking client-provided contracts and financial documents for consistency. Interviews with stakeholders, such as clients, executives, or third parties, complement document review, enabling lawyers to detect hidden risks such as undisclosed related-party transactions or equity disputes. Several participants noted that early interviews often determine the scope of due diligence and guide focus areas. For high-stakes transactions like Initial Public Offerings (IPOs) or major investments, on-site verification is employed to confirm operational realities and asset values, especially when discrepancies cannot be detected remotely. Among the participants, 12 out of 18 emphasized that these site visits are critical for regulatory compliance and risk mitigation, providing firsthand evidence that cannot be fully captured through documents or interviews alone.

While verification tasks share common modes, their organization varies substantially across non-litigation practices. Lawyers stressed that workflows differ by business type, transaction complexity, and client objectives. Therefore, steps cannot be easily standardized. For example, IPO projects prioritize regulatory compliance and disclosure, Mergers and Acquisitions (M\&A) and Private Equity (PE) transactions focus on transaction feasibility and risk allocation, and ongoing advisory work centers on problem decomposition and document preparation. Despite these differences, an overarching logic is observed: legal reasoning guides the workflow, starting with identifying the relevant legal framework, proceeding through fact verification, iteratively adjusted as new information emerges, and culminating in the legal deliverable. This iterative sequencing, combined with the integration of multiple information sources and verification modes, highlights the potential for GenAI tools to support routine data collection and preliminary consistency verification, while human judgment remains central to decision-making, interpretation, and accountability.

\subsubsection{Complexities in Fact Verification.}
\label{4.1.2} 
Fact verification involves multiple layers of difficulty for lawyers, who must reconcile fragmented information sources, evaluate the credibility of inconsistent disclosures, bridge gaps in technical knowledge, and coordinate with multiple stakeholders. Together, these factors make verification a time-consuming and cognitively demanding part of their work.

14 out of 18 lawyers highlighted that fragmentation across public information platforms such as corporate registries, court databases, credit reporting systems, and regulatory disclosure portals adds substantial workflow overhead. As P7 explained, \textit{“There is no unified entry point, and some platforms don’t update on time, so you can’t tell if missing data means it doesn’t exist or just hasn’t been published.”}
Moreover, platform fragmentation exacerbates inconsistencies or outdatedness in publicly available information, since different platforms host varying versions or timelines of data, necessitating manual cross-verification (P4, P6, and P9). Additionally, some information is only accessible via client disclosure, requiring interviews or on-site investigations (P2, P3, and P11). 

Lawyers further highlighted that determining the adequacy and credibility of evidence is among the most complex aspects of fact verification. This complexity stems from fragmented sources and the potential bias or incompleteness of self-disclosed information. As P6 noted, \textit{“The hardest part is that you can’t just rely on what the company tells you. Some shareholders may conceal details like proxy holdings, and we need to judge if there’s something hidden behind their statements.”} Among the participants, 13 out of 18 reported that even publicly disclosed information may be curated to present a favorable image, requiring cross-verification with financial statements or third-party records. Detecting inconsistencies and identifying “red flags” relies heavily on professional judgment and accumulated experience (P2 and P9).

Verification also frequently extends beyond legal expertise into specialized domains such as biotechnology, engineering, or data compliance. Lawyers emphasized that even experienced practitioners cannot master all legal subfields or associated technical knowledge. P9 remarked, \textit{“The legal industry is highly segmented. Each lawyer typically focuses deeply on one or two areas.”} Among the participants, 12 out of 18 reported that unfamiliarity with technical or regional regulations significantly slowed verification and increased the risk of overlooking compliance points. For example, P11 highlighted regional differences in labor law implementation affecting maternity allowance eligibility, and P5 noted the high cognitive load when encountering technical fields for the first time. P7 further warned that an incomplete understanding could lead to overlooking critical compliance requirements, particularly in intellectual property and advertising cases.

Beyond information hurdles and expertise gaps, coordination across multiple parties constitutes an additional barrier. Among the participants, 11 out of 18 reported. Many verification tasks rely on collaboration with clients, counterparties, underwriters, or accountants. Scheduling interviews is often inefficient, especially when stakeholders are uncooperative. P2, P3, and P16 noted that information from clients or target companies can be inaccurate or incomplete. P8 described the difficulty of arranging shareholder interviews, \textit{“Some former shareholders are hard to reach and not very cooperative.”} 

Overall, the complexities of fact verification arise from the interplay of fragmented information sources, ambiguous disclosures, gaps in specific domain knowledge, and the need to coordinate with multiple stakeholders. These overlapping challenges mean that verification is not a straightforward process of gathering facts but a continuous effort to interpret incomplete, inconsistent, or inaccessible information while managing dependencies across different actors and systems. The result is a demanding, iterative workflow in which lawyers must frequently revisit earlier steps, refine their judgments, and reconcile inputs from diverse sources.

\subsubsection{GenAI Can Reduce Uncertainty and Burden in Fact Verification.}
\label{4.1.3} 
Lawyers reported that GenAI primarily serves a supportive role in the fact verification process, helping them navigate the complexities described in Section \ref{4.1.2}. Rather than replacing professional judgment, GenAI is integrated into the existing workflows to alleviate cognitive load, streamline preparatory work, and assist in early-stage fact verification. In practice, these applications cluster around three interrelated levels aligned with the fact verification: providing cognitive support in unfamiliar domains, assisting in the generation and refinement of routine legal documents, and structuring the early stages of fact verification and information search.

\textbf{Cognitive support for unfamiliar domains.}
One major complexity identified in Section \ref{4.1.2} is the need to verify facts in industries beyond a lawyer’s core expertise, such as biotechnology, pharmaceuticals, engineering, or data compliance. To address these knowledge gaps, lawyers use GenAI for orientation rather than authoritative advice, helping them frame the problem space before conducting rigorous legal research. P9 explained, \textit{“If I’m assigned to a new area, I might first ask AI for a broad overview, then use that to guide my own research.”}
Several participants emphasized using GenAI to quickly understand regulatory landscapes and identify risk points, especially for due diligence in unfamiliar sectors. P13 described, \textit{“If the target is in an unfamiliar industry, such as pharmaceuticals, I use AI to check due diligence risk points, regulatory focuses, and to clarify the direction for deeper investigation.”} Similarly, P14 added, \textit{“I tell AI the specific background and requirements, and it generates a checklist to ensure no key steps are missed. AI gives me a clear direction and helps me quickly organize my thoughts.”}

GenAI also supports terminology clarification and company background research, reducing the time spent on initial exploration. P1, P2, and P3 reported using tools like ChatGPT or DeepSeek to obtain summaries, definitions, and initial references, which were always followed by professional verification. As P18 summarized this benefit, \textit{“As a lawyer, I have to read a lot of materials. If I rely only on myself, it’s time-consuming. AI helps extract key information efficiently.” } While all outputs require verification, participants agreed that GenAI reduces the cognitive overhead of entering new knowledge domains.

\textbf{Support for generating and refining routine legal documents.} A second integration point concerns the heavy document load surrounding fact verification, such as due diligence templates, draft opinions, background summaries, multilingual communications, and more. 10 participants reported that it reduced the time and cognitive effort associated with drafting routine documents.
As P9 observed, \textit{“The biggest value is reducing time costs. Many tasks can be generated as a draft by AI, so our work shifts from ‘starting from scratch’ to ‘editing and refining a draft,’ which greatly improves efficiency.”}

P1 described delegating routine drafting, \textit{“AI saves me time by drafting opinion letters once I provide basic transaction details.”} Similarly, P2 emphasized that GenAI accelerates workflow setup, \textit{“I can give AI the key facts and quickly get a draft legal opinion.”} Lawyers highlighted the efficiency gains from this shift, as P7 noted, \textit{“Our job used to start from zero, but now it starts from a draft.”}

Beyond first drafts, GenAI supports the customization of templates for new projects, summarization of long disclosures and translation and polishing (P2, P3 and P16). P2 described customizing templates, \textit{“We provide a template from another project, specify current project requirements, and let AI adjust the content to fit the current project, rather than just doing text edits.”} P3 highlighted long document summarization, \textit{“Upload long files to AI, generate core content summaries, saving reading time.”} P16 stated, \textit{“For instance, I might draft an email in English, then use ChatGPT to polish it.”} These applications do not conduct verification per se, but they reduce the preparatory burden that precedes verification, creating more time and cognitive bandwidth for judgment-dependent evaluation.

Finally, GenAI supported peripheral tasks like presentation preparation, email drafting, or even multilingual introductions for client interactions, as P14 noted \textit{“When receiving international guests, AI helps generate phonetic pronunciations and profile summaries to facilitate communication.”} 

However, participants reaffirmed that all AI-generated documents demand careful review and adaptation to meet legal standards. GenAI assists with pace and structure, not with substituting professional interpretation.

\textbf{Structuring early-stage fact verification and search processes.}
Lawyers also reported leveraging GenAI to structure early verification steps, particularly search formulation, preliminary risk spotting, and initial contract review. These applications address the complexity of fragmented information sources and the iterative nature of verification identified in Section \ref{4.1.2}, by helping lawyers navigate fragmented platforms and identify where to focus scrutiny. 

One common practice involves using GenAI to generate keywords for factual searches, thereby improving retrieval efficiency across multiple platforms. P9 explained, \textit{“Before every search, I ask AI to suggest keywords for the problem I’m facing. Having the keywords makes later searches much smoother.”} Similarly, GenAI was applied to legal preliminary case retrieval and statutory research, providing summaries and candidate references that lawyers then validate manually (P1, P5, and P12).

For contract review, lawyers described GenAI as a tool for idea generation and structured risk scanning. P9 noted, \textit{“AI can provide ideas, such as key points for contract review, though the accuracy is uncertain and still needs verification.”} P11 highlighted the value of specialized systems such as Maka Contract, \textit{“AI can help me identify risk points I might overlook in a single review, compensating for blind spots and improving the comprehensiveness of contract verification.”} Similarly, P3 shared, \textit{“AI provides initial risk directions for contracts, and then we manually refine them to improve efficiency.”}
P10 described asking AI to \textit{“examine each clause, provide risk alerts and revision suggestions, and cite the relevant legal basis,”} which lawyers then refine. Even when accuracy is imperfect, these early structures help lawyers triage attention and reduce oversight risk in routine reviews. P1 added that specialized systems allow configuration by transaction roles and neutrality settings before producing outputs that lawyers manually validate, \textit{“After human review, the major points are basically covered.”}

Some participants also used GenAI for due diligence verification and company background verification, particularly in generating structured outputs. P15 noted, \textit{“Some AI tools can perform basic company research and return results in Word format.”} P18 recalled prior experience, \textit{“I’ve used AI to assist in checking litigation records when drafting due diligence reports.”} Across these applications, GenAI does not confirm factual adequacy but helps organize and prioritize verification tasks in environments where data is dispersed and inconsistent.

Despite its growing integration, lawyers consistently emphasized that GenAI cannot determine factual credibility or completeness. As P12 warned, relying on general tools without professional verification \textit{“could cause huge risks.”} P3 similarly stressed that \textit{“AI can’t understand the full context behind legal decisions.”} As a result, most participants framed GenAI as a burden-reducing tool that is valuable for orientation, handling document, and structuring tasks, but not yet capable of replacing professional judgment in evaluating whether facts are adequate, reliable, or strategically curated. As P13 noted, \textit{“I won’t directly use AI’s outputs, whether it’s legal citations or due diligence frameworks, I have to adapt everything to the project’s specifics.”} Rather than replacing existing fact verification procedures, GenAI is incorporated as an upstream aid that accelerates information scanning, highlights potential inconsistencies, and surfaces preliminary directions for manual confirmation. However, the very factors that make GenAI useful also introduce new points of vulnerability, which complicate its integration into high-stakes fact verification. The following section examines the challenges and risks that arise from this tension.

\subsection{Challenges Shaping GenAI Adoption}
\label{4.2}
Interviews with 18 lawyers reveal that despite the growing visibility of GenAI in legal practice, its integration into fact verification processes remains cautious and tightly bounded. Rather than treating technological limitations as isolated issues, participants consistently described how the nature of fact verification itself heightens the stakes of GenAI use. Fact verification requires zero tolerance for error, strict control over sensitive information, and accountability for every conclusion drawn, all of which shape when, how, and to what extent lawyers can rely on GenAI. Across interviews, two interrelated challenges emerged. First, the professional responsibility lawyers bear for the accuracy and defensibility of their work amplifies the risks associated with GenAI errors. Second, concerns about information control, including confidentiality, privacy, and the protection of client data, place structural limits on how GenAI can be integrated into verification workflows.

\subsubsection{The Demands of Legal Accountability Magnify the Impact of GenAI Errors.}
\label{4.2.1}
15 out of 18 lawyers agreed that GenAI’s inaccuracies are not problematic merely because they exist, but because legal accountability frameworks substantially magnify the consequences of even small mistakes. In legal fact verification, every factual reference, whether a regulation, clause, or interpretation, must be verifiable, defensible, and ultimately signed off by a human lawyer. This requirement transforms GenAI’s well-known failure modes into significant professional risks. As a result, the threshold for acceptable error is far lower than what exists in other domains.

Participants frequently described situations in which GenAI produced fabricated, outdated, or misleading legal information, and emphasized that once such content enters a draft, it becomes the lawyer’s own liability (P8, P10 and P11). Several encountered hallucinated legal references that appeared unusually credible. As P6 explained, \textit{“AI-generated misinformation is often logically structured and appears well-supported, making it highly convincing. However, subsequent verification frequently reveals that there are no corresponding original texts or sources to back its claims.”} Because lawyers are accountable for the accuracy of every element of their work, regardless of whether it originated from GenAI, these kinds of outputs require exhaustive review. 

This dynamic also reshapes lawyers’ workflows. Instead of accelerating fact verification, GenAI often imposes additional verification burdens. Participants noted that when GenAI presents information confidently, they must devote significant time to tracing each citation back to its source. As P13 described, \textit{“I would first look at the results given by GenAI, and then find the original links based on the data it cited to check whether GenAI’s reasoning is in line with the description in the original materials. For example, when GenAI interprets a certain regulation or policy, I must trace back to the official original text to confirm there is no deviation.”} Such backtracking, they explained, can consume more time than conducting verification manually.

The high stakes associated with fact verification further amplify the impact of GenAI’s errors. Lawyers pointed out that the models frequently struggle with multi-step reasoning or complex legal scenarios. When errors are subtle or embedded within otherwise coherent outputs, they can escape initial detection and surface only during later stages of review, by which point the cost of correction is significantly higher. As P11 noted, \textit{“The difficulty in verifying regulations lies in the fact that the publicly available information is too extensive and redundant. It’s very easy to come across false regulations and ordinances on the Internet, and one can easily be misled. Therefore, it takes a lot of time to identify the authenticity of the information sources.”}

These accuracy issues intersect directly with professional responsibility. Lawyers repeatedly emphasized that GenAI cannot be trusted for final verification because accountability ultimately lies with them, not the tool. As P18 stated, \textit{“(GenAI) is unlikely to completely replace (humans). Ultimately, verification still needs to be done by people.”} Similarly, P11 stressed that \textit{“AI can assist with first drafts or provide initial suggestions, but every clause must be manually verified to ensure it complies with legal standards.”} This recognition reinforces the profession’s zero-tolerance stance toward unverified outputs. 

As a result, GenAI can generate large amounts of information quickly, but the heightened scrutiny required to validate its outputs often offsets these benefits. Instead of functioning as a dependable verifier, GenAI becomes an additional object of verification that lawyers must carefully audit to avoid professional and legal repercussions. In this context, the demands of legal accountability magnify the practical impact of AI errors, limiting the extent to which lawyers can integrate GenAI into core verification tasks.

\subsubsection{Concerns About Information Control Limit How Lawyers Can Use GenAI}
\label{4.2.2}
While accuracy concerns constrain how GenAI outputs are used, questions of information control shape whether sensitive materials can be entered into GenAI systems at all. 14 lawyers emphasized that fact verification routinely involves highly confidential documents, such as transaction records, personal identification materials, internal correspondence, and strategic evaluation, where even minor disclosure could trigger legal, commercial, or reputational harm. Because lawyers are responsible not only for factual accuracy but also for safeguarding client information, the perceived opacity of GenAI significantly limits their adoption in verification workflows.

Participants consistently expressed apprehension about uploading sensitive materials to external GenAI platforms. As P13 warned, \textit{“The main concern is data privacy leakage... If we upload this data to the AI platform, we’re worried that it might be identified and leaked, especially the core transaction information. Once leaked, it could affect the progress of the project and even trigger legal risks.”} This risk calculus is particularly acute in fact verification, where the documents that require verifying are often the most sensitive in a case. Similarly, P1 underscored the personal burden lawyers bear in protecting client data, \textit{“Anonymization needs to be handled by ourselves. We are extremely concerned about data security issues, because much of this data involves corporate privacy.”}

To maintain control over sensitive information, some firms experimented with internal or on-premise GenAI tools, often coupled with automated redaction or desensitization features. Yet these tools still require manual oversight to satisfy confidentiality obligations. P10 noted, \textit{“Our desensitization tool can remove names and IDs with one click, but it’s not precise enough—manual checks are still needed.”} Others, like P11, described taking on additional labor to ensure compliance, \textit{“Carry out data desensitization in advance. For particularly sensitive contracts, manually hide the client’s core privacy information, such as trade secrets and personal information. Then upload the contracts to AI tool, reducing the risk of leakage at the source.”} These mitigations reduce exposure but do not eliminate the underlying uncertainty about where data goes and who can access it.

Importantly, concerns about information control also intersect with professional responsibility. Lawyers stressed that the decision to use GenAI is not theirs alone. It must align with the client’s confidentiality expectations. As P17 explained, \textit{“Respect the client’s wishes. If the client has high confidentiality requirements and clearly prohibits the use of GenAI, regardless of the actual confidentiality capabilities of GenAI, we must follow the client’s request. If the client deems the confidentiality requirements to be low and agrees to use GenAI, then we can proceed with the subsequent operations.”} This relationship means that GenAI integration is shaped as much by ethical and contractual obligations as by technology.

These constraints also reflect a deeper trust boundary. Many participants highlighted that the opacity of GenAI about where data travels, how it is stored, how it is used for model training prevents them from treating GenAI as reliable partners in fact verification. This professional caution echoes their concerns about accuracy, just as lawyers must review all AI-generated outputs to avoid liability, so they must also tightly control the flow of information into GenAI to avoid breaching client confidentiality.

Taken together, the core obstacles to GenAI adoption in fact verification are not merely technical shortcomings but the misalignment between current GenAI and the epistemic, ethical, and legal conditions under which lawyers must work. Because fact verification requires both high-fidelity information and strict control over how that information circulates, any technology that introduces uncertainty, whether through hallucinated outputs or opaque data-handling practices, heightens the risks lawyers are accountable for. As a result, GenAI is used selectively and cautiously, confined to tightly bounded tasks where its limitations can be contained. These frictions point that lawyers do not simply need GenAI that is “more accurate” or “more secure,” but systems that can integrate into their verification responsibilities without reshaping the very standards they are obligated to uphold. The next section explores how lawyers imagine future forms of human–AI collaboration that could better meet these demands.

\subsection{Envisioning Future Human-AI Collaboration}
\label{4.3}
Participants’ imaginations did not portray GenAI as simply becoming more capable, but as transforming how legal fact verification should be organized to maintain accuracy, accountability, and professional autonomy. Rather than accepting current limitations, lawyers articulated what a workable partnership with GenAI would require: (1) automation as a strategy for redistributing cognitive labor, (2) transparency as a mechanism for reinstating professional trust and (3) specialization as a path to align GenAI with legal epistemic norms.

\subsubsection{Automation as a Strategy for Redistributing Cognitive Labor.}
\label{4.3.1}
Across interviews, automation emerged as an intentional strategy for reallocating cognitive effort away from repetitive preparation work and toward interpretive, judgment-dependent tasks. 9 out of 18 participants explicitly envisioned AI systems that take over labor-intensive front-end activities, such as processing interview recordings, extracting compliance-related information, or generating structured working papers, so that lawyers can concentrate on the reasoning steps that machines cannot reliably perform.

Participants emphasized that the value of automation lies in shifting rather than removing human involvement. P8 illustrated this orientation when describing the desired workflow, \textit{“If AI tools can automatically organize interview recordings, extract key information, and generate standardized working papers, it can significantly reduce the manual workload.”} In this framing, automation is a mechanism for clearing the cognitive space required for higher-order analysis.

This redistributive view also shaped expectations for advanced report generation. 3 lawyers (P1, P3 and P12) described one-click production of due diligence reports in which GenAI aggregates corporate information, organizes evidence, and produces a first-pass draft that they would subsequently refine. P12 articulated the underlying logic by invoking an established professional model, \textit{“In our industry, we call this model ‘human-in-the-loop’. I think it’s more suitable for scenarios like the legal field, which is highly professional and has a low tolerance for errors.” } Here, automation is explicitly framed as augmenting human judgment rather than replacing it.

Several participants extended this logic by imagining AI agents that sustain continuity across tools and tasks through memory and adaptive learning. P11 explained, \textit{“If there were a butler AI on the computer that could record my usage data and work preferences and synchronize them to new tools, it would save me the hassle of repeated settings.” }Rather than autonomy, what lawyers sought was an intelligent infrastructure that supports long-term workflow stability, allowing them to invest more cognitive effort into legal reasoning and less into operational overhead.

Collectively, these accounts position automation as a means of redistributing cognitive labor. AI systems handle structure, standardization, and extraction, while lawyers retain responsibility for interpretation, judgment, and final accountability. This framing reflects not only efficiency concerns but also an effort to preserve professional expertise in the face of increasingly capable GenAI.

\subsubsection{Transparency as a Mechanism for Reinstating Professional Trust.}
\label{4.3.2}
7 participants repeatedly emphasized that transparency is not merely a technical enhancement but the precondition for allowing GenAI to participate meaningfully in legal fact verification. After encountering hallucinations, incomplete reasoning, and unverifiable claims in current systems, lawyers argued that AI-generated information becomes professionally usable only when its origins and inferential steps are exposed in ways that match the standards of legal accountability. Transparency, in this sense, functions as the mechanism through which trust—eroded by uncertain accuracy—can be re-established.

A central expectation was that every factual statement produced by GenAI must be traceable to an authoritative and inspectable source. Lawyers stressed that without clear provenance, such as timestamps, screenshots, or links to official registries, AI outputs cannot be integrated into verification workflows. As P18 put it, \textit{“When AI outputs shareholder information, it is necessary to indicate that it is from a ‘screenshot of the Enterprise Credit Information Publicity System on X, X, 2024’ and attach a link or screenshot.”} Traceability restores the lawyer’s ability to conduct independent verification, which remains non-negotiable even in AI-supported workflows.

Participants also called for transparency that mirrors the structure of legal reasoning rather than presenting information as undifferentiated text. Many envisioned outputs that unfold through layered, navigable formats, such as mind maps, issue trees, or structured reports, that lay out the relationship between legal issues, factual findings, and underlying legal bases. P7 described this expectation, \textit{“I hope that AI can output reports or mind maps with footnotes. The mind map can be presented in layers according to ‘legal issues-corresponding situations-legal basis.’ Each conclusion should be marked with its source, such as a specific law or a link to an official website, and clicking on it can display the full text.”} Such formats allow lawyers to inspect, question, and revise the reasoning path itself.

Several participants further argued that transparency should help them prioritize verification efforts by signaling how reliable different claims are. P14 articulated this clearly, \textit{“For example, if it can judge based on the source and mark ‘information from the official website, credibility 99\%’, it will be more valuable for reference.”} These indicators guide where verification should be concentrated, allowing lawyers to triage the workload in a process where time is limited but error tolerance is near zero.

Across interviews, transparency was also understood as an interactional requirement rather than a static documentation practice. Participants imagined interfaces that allow them to click through citations, jump to original documents, inspect AI reasoning step by step, and reconcile summaries with full-text evidence within a single workflow. This expectation signals that transparency is meaningful only if it is manipulable.

In conclusion, these expectations illustrate that transparency is the pathway through which professional trust can be rebuilt. It is not because transparency ensures correctness, but because it reinstates the lawyer’s capacity to govern AI-generated information. By making sourcing explicit, reasoning inspectable, credibility legible, and verification interactive, transparency aligns GenAI with the evidentiary standards that legal fact verification demands.

\subsubsection{Specialization as a Path to Align GenAI with Legal Epistemic Norms.}
\label{4.3.3}
Lawyers emphasized that the shortcomings of GenAI stem not only from errors in output but from a deeper epistemic misalignment. It refers to that these models do not reason according to the criteria lawyers use to determine whether a fact is legally valid. Participants therefore framed specialization as the mechanism by which GenAI could adopt the epistemic norms embedded in legal practice. This alignment would require specialization along two dimensions. They are internal reasoning, which refers to how the model determines what constitutes a valid legal fact, and operational representation, which involves how the model presents and handles information in forms that facilitate legal verification.

The first dimension concerns the model’s internal reasoning logic, which must reflect legal distinctions such as authority hierarchies, source validity, and statutory currency. As P5 noted, \textit{“I hope there can be a dedicated AI tool for legal fact verification, such as a tool that can accurately identify the validity of legal provisions and automatically exclude fictional cases.”} What P5 demands is that the underlying reasoning of GenAI embodies the normative verification criteria lawyers routinely apply, whether a provision is in force and whether a claim is grounded in a recognized legal source. 

The second dimension involves operational representations, where specialization allows GenAI to express information using domain-specific visual and structural forms that lawyers rely on to interpret legal relationships. This was not framed as a formatting preference but as a way to align GenAI with the representational practices that structure legal inquiry. P13 illustrated, \textit{“For example, it can automatically organize the shareholder structure into a table and even generate an equity structure diagram, reducing the time spent on manual tabulation and drawing. After all, drawing equity structure diagrams is very time-consuming nowadays.”} GenAI that can generate legally meaningful diagrams or tables does not merely offer convenience. It demonstrates an understanding of the epistemic conventions governing the verification of legal facts, such as how relationships like ownership, control, or liability must be visualized to facilitate scrutiny.

These perspectives reveal why specialization was seen as essential. Rather than improving accuracy in a generic sense, participants envisioned specialized AI systems that recognize legally salient distinctions, reason with domain-appropriate criteria, and present information in forms that make verification intelligible. In their view, only through such multi-layered specialization could GenAI become a trustworthy collaborator in legal tasks.
\section{Discussion}

\subsection{Enhancing Efficiency Through Structured and Epistemically Aligned Automation}

Efficiency in legal fact verification cannot be understood as simply accelerating routine tasks. The findings in Section \ref{4.3.1} reveal that what lawyers seek is a form of automation that strengthens the evidentiary infrastructure of their work rather than replacing human reasoning. Because fact verification in non-litigation practice involves reconstructing a defensible account of “what is true” across fragmented registries, client disclosures, interviews, and on-site observations, efficiency gains depend on whether GenAI can reduce uncertainty at the earliest stages where facts are gathered, organized, and stabilized. In this process, lawyers carry full accountability for every factual claim they endorse. As a result, automation is acceptable only when it reduces ambiguity without altering the chain of evidentiary interpretation.

Section \ref{4.1.3} shows that the heaviest cognitive burden lies not in legal reasoning itself but in the upstream work of transforming unstructured information into verifiable elements. Tasks such as extracting shareholder changes from interview notes, comparing multiple versions of corporate records, or building chronological transaction tables require continuous attention to consistency and timestamp alignment. These preparatory steps determine which details can be safely carried into risk evaluation, disclosures, or due diligence deliverables. It reveals that while automation in many fields offloads repetitive tasks, in law, similar offloading risks inadvertently shifting epistemic responsibility. AI-generated summaries often appear convincing but provide little visibility into the selection or interpretation of information \cite{sun2024ai}. Once such content enters a working paper, verifying it requires retracing the system’s steps, often consuming more time than manual work \cite{sahoh2023role}. Efficiency collapses not only due to inherent GenAI inaccuracy, but more fundamentally because its opacity obscures the source of error \cite{hutson2021opacity}, thereby disrupting the evidentiary chain that lawyers must preserve, which in turn suggests automation's true value lies not in removing simple work but in its potential to stabilize the compromised factual substrate.

The inability to delegate extraction and organization due to accountability needs means that efficiency is achieved only when GenAI acts as an infrastructural intermediary, which directly reduces the cognitive load caused by fragmented data environments by producing structured intermediate artifacts with explicit provenance. This enables lawyers to begin their reasoning from a more reliable evidentiary baseline. Structured GenAI that can automatically align data across sources, flag timestamp discrepancies, or generate audit trails directly makes it possible for lawyers to shift their cognitive effort from reconstructing factual coherence to evaluating legal implications \cite{mora2021traceability}. Meaningful automation depends on systems that first act as stabilizing forces for data fidelity, ensuring that the material lawyers interpret remains trustworthy. Only then can the boundaries of human legal responsibility be clearly defined. Ultimately, this reframes efficiency as an epistemic design problem, demanding that AI systems be engineered to support the evidentiary commitments of legal practice rather than merely its workflow timelines.

\subsection{Preserving Expertise within AI-Supported Legal Fact Verification}

While GenAI offers opportunities to alleviate cognitive burden in fact verification, our findings in Section \ref{4.1.2} show that professional expertise remains essential, particularly when lawyers must navigate uncertainty, reconcile fragmented evidence, and make defensible judgments. And the most demanding parts of verification are not mechanical but interpretive, such as identifying omissions in client disclosures, recognizing strategically curated information, or detecting inconsistencies across platforms. Increasing automation may narrow the hands-on exposure through which early-career lawyers build familiarity with evidentiary nuances.

Junior practitioners typically develop judgment by conducting the labor-intensive groundwork of reviewing documents, cross-checking registries, and piecing together partial information \cite{otto1989identifying}. These activities expose them to the texture of real cases, such as the signs of unreliable narratives, indicators of regulatory risk, or subtle discrepancies that warrant further investigation. When AI systems begin to streamline these early-stage tasks, they risk removing the very conditions through which evidentiary sensitivity is cultivated. 

These dynamics reshape the role of AI literacy in legal practice, which must help lawyers preserve the capacity for independent judgment within AI-supported workflows. Our findings suggest three competencies that are especially relevant. First, lawyers need the ability to interrogate the provenance and reasoning behind GenAI outputs, understanding where the model’s conclusions originate and which evidentiary assumptions they embed. This matters because errors in GenAI reasoning are often subtle, especially when hallucinated facts appear coherent or well-structured \cite{li2025simulated}. Second, lawyers must learn to recognize which steps in the verification process remain irreducibly human, such as evaluating credibility, weighing conflicting information, or determining whether evidence is sufficient to meet professional accountability standards. Finally, lawyers need strategies for calibrating when to rely on GenAI support and when manual verification is necessary to maintain defensibility, which trains lawyers in process orchestration.

By framing AI literacy around these competencies, the goal is to maintain the depth of expertise needed to use GenAI safely and effectively \cite{lewis2024all}. The challenge ahead is to integrate AI systems in ways that reduce unnecessary cognitive load without narrowing the experiential space through which legal judgment is developed. Sustaining this balance is central to ensuring that increases in efficiency do not come at the expense of the professional reasoning that underpins trustworthy legal practice.

\subsection{Systemic Risks Introduced by GenAI in Fact Verification}
GenAI does more than introduce occasional errors into fact verification. It reshapes the epistemic conditions under which legal facts are constructed, interpreted, and justified.
The following sections will detail these systemic challenges, demonstrating how they fundamentally alter the relationships in legal fact verification. Furthermore, based on our findings, GenAI must be treated as an actor that remains auditable, constrained, and aligned with the evidentiary commitments of legal practice.

\subsubsection{Epistemic Ambiguity in AI-Mediated Evidence Formation.}

A central insight from our findings is that AI systems do not merely accelerate fact verification, a benefit often neutralized by high verification costs. Instead, they subtly reshape the evidentiary landscape by producing new forms of ambiguity that lawyers must now anticipate and manage. This stems from the unstable epistemic status of AI-generated content \cite{burrell2016machine}. Unlike traditional documentary evidence, such as contracts, regulatory filings, or physical invoices, which carries a clear provenance and an auditable trail, AI-generated summaries, extracted facts, or suggested inferences are neither raw data nor fully validated legal facts. Such AI-mediated content may inadvertently shape how lawyers initially frame specific evidentiary connections by foregrounding specific connections while actively obscuring others \cite{yuan2023contextualizing}. This creates a significant risk that early-stage interpretations acquire unwarranted momentum simply because they are algorithmically amplified. Consequently, the initial pass through AI outputs is transformed from a neutral shortcut into a potential intervention in how facts are conceptualized and prioritized for legal strategy.

The ambiguity is heightened by the difficulty of reconstructing how the system arrived at specific suggestions. Even small inaccuracies in an AI-generated summary could propagate through downstream reasoning, especially when working with tight deadlines or large volumes of disclosure material. Rather than helping them see more, GenAI can narrow their attention to what the system presents as salient, introducing a new class of blind spots that differ fundamentally from human oversight errors \cite{cummings2017automation}. These are not omissions arising from fatigue or limited time, but gaps structurally produced by the AI’s model architecture, training data, and probabilistic inference patterns.

This emerging form of evidentiary ambiguity complicates one of the core elements of legal fact verification, which is the ability to justify not only what was concluded but how it was reached. As our findings in Section \ref{4.2.1} suggest, lawyers need to spend additional effort identifying which parts of an AI output were reliable, increasing the burden of checking underlying evidence. AI-generated content creates uncertainty about what counts as the starting point of  fact verification, forcing lawyers to develop new practices for verifying whether automated outputs distort the evidentiary field. The challenge is maintaining clarity about the boundaries between evidence, interpretation, and AI-mediated inference. 

To address this evidentiary ambiguity, systems should foreground the epistemic status of AI-generated outputs rather than masking them as authoritative or complete. This includes explicitly signaling when content is inferred rather than extracted, marking low-confidence or model-constructed connections, and allowing lawyers to see distinctions between verbatim source text and AI-generated paraphrase or synthesis. Such forms of epistemic transparency do not eliminate ambiguity but reposition it as an inspectable property of the system, enabling lawyers to scrutinize how AI-mediated inferences may be shaping the early contours of a case. It can help lawyers reassert control over the interpretive starting points that anchor legal fact verification by making ambiguity visible and traceable.

\subsubsection{Integrity of Transformed Evidence.}
\label{5.3.2}

The fragility of evidence integrity constitutes a second ethical challenge stemming from GenAI's intervention in factual material. As our findings illustrate, lawyers rely on the stability and traceability of documentary evidence so they can interrogate origin, context, and limitations. AI-assisted workflows disrupt this stability by inserting an additional layer of transformation between the raw material and the lawyer’s evaluation \cite{kudina2019ethics}. Even when systems do not hallucinate, they may reformat, paraphrase, or recontextualize information in ways that subtly shift its evidentiary meaning. This transformation is not through intentional misrepresentation but through the system’s optimization for brevity or coherence \cite{caraban201923}.

This transformation is ethically consequential because legal reasoning depends on a chain of custody not only for documents but for meaning itself. When GenAI intermediates this chain, the boundaries between original evidence and derivative representation become blurred. This creates a risk that AI-produced paraphrases sometimes shifted emphasis or omitted context, which could influence what appeared salient in early evaluation.
The opacity of model behavior compounds this risk. AI systems provide no transparent rationale for why certain portions of a document were condensed or why specific terms were highlighted. This makes error detection more labor-intensive and shifts part of the evidentiary burden from examining the content to examining the transformation process.

Therefore, systems should make visible the transformations applied to source material, highlighting what was condensed, omitted, or reframed. This is essential so that lawyers can audit not only the output but also the process by which the output was produced. Interfaces should expose transformations as inspectable layers to maintain the integrity of legal evidence.

\subsubsection{Misaligned Control and Accountability in AI-Assisted Reasoning.}

A third ethical concern is the misalignment between where legal responsibility is assigned and where control is practically exercised. As discussed in Section \ref{4.3}, lawyers are required to maintain authorship over the factual and interpretive steps that inform their conclusions. GenAI complicates this obligation not through opacity alone, as addressed in Section \ref{5.3.2}, but by inserting AI-generated structures at stages where professional accountability norms assume direct human authorship. Once such structures shape the downstream organization of evidence, it becomes difficult for lawyers to ensure that the reasoning embedded in these intermediate steps corresponds to their own standards of justification.

This creates that lawyers remain fully liable for the representations they endorse even when they have limited control over the interpretive scaffolding that influenced those representations. The problem is not merely that system outputs are imperfect, but that they can steer reasoning in ways that are difficult to revisit without reconstructing an earlier evidentiary state \cite{novelli2024accountability}. Under these conditions, AI-produced structures could influence their reasoning before they had fully examined the underlying evidence, complicating how they justified later decisions.

To address this misalignment, systems should be designed to ensure that interpretive authority and the opportunity to exercise it remain with the lawyer at the stages where liability is later attributed. Design should constrain the degree to which early system-generated structures impose a pre-formatted structure on the evidentiary record. Useful interventions include requiring explicit human confirmation before adopting AI-generated organizational schemas, offering reversible transformations that allow lawyers to return to pre-AI states, and constraining default inferences in high-liability tasks. These measures help preserve the alignment between control and accountability, supporting responsible reliance without narrowing professional agency.

\subsection{Limitations and Future Work}
As with most interview-based research, our study has limitations that should be considered when interpreting the findings. Because the data relies on practitioners’ self-reported accounts, participants’ descriptions of their verification routines and GenAI use may diverge from actual practices, particularly in fast-paced non-litigation workflows. Although the interviews encouraged concrete examples and detailed reflection, discrepancies between articulated strategies and enacted behavior may persist. Future research could complement this work with observational studies to examine how G enAI-mediated verification unfolds in practice. 

In addition, while our discussion engages with prior work on GenAI in high-stakes domains, the analysis does not provide a comprehensive engagement with adjacent literatures such as health AI or other high-stakes professional domains. This limits the extent to which our findings are positioned within broader cross-domain debates in HCI. Future research could build on this work by conducting comparative or integrative analyses across professional settings to examine shared and divergent expectations for GenAI-supported verification.

Finally, the study focuses on non-litigation. While this perspective offers rich insight into sensemaking, documentation, and risk evaluation in everyday legal work, future research could extend to litigation to examine whether AI-mediated verification practices differ across domains.
\section{Conclusion}
\label{6}
In this study, we conducted a qualitative analysis of interviews with 18 legal professionals to explore their experiences, attitudes, and expectations regarding GenAI integration in legal fact verification. 
Our findings reveal that GenAI not only change the pace of fact verification but also influence how lawyers interpret information, form judgments, and manage responsibility within their workflows. Participants described both the benefits of reduced manual burden and the risks that arise when AI-generated content shapes early understanding of the fact or interrupts opportunities for developing professional intuition.
These insights highlight the need for auditable AI systems that support transparency, preserve human oversight at critical decision points, and maintain conditions that allow legal expertise to grow. By grounding these implications in real practitioner experiences, our study offers guidance for designing AI tools that can be adopted responsibly in everyday legal work. More broadly, the work contributes to HCI research on expert–AI collaboration by showing how efficiency gains must be balanced with the practices that enable reliable professional reasoning.

\begin{acks}
This work is funded in part by the HKUST Start-up Fund (R9911), Theme-based Research Scheme grant (T45-205/21-N), the InnoHK initiative of the Innovation and Technology Commission of the Hong Kong Special Administrative Region Government, and the research funding under HKUST-DXM AI for Finance Joint Laboratory (DXM25EG01).
\end{acks}

\bibliographystyle{ACM-Reference-Format}
\bibliography{references}

\appendix

\end{document}